# Enhancing Binomial and Trinomial Equity Option Pricing Models

**Yong Shin Kim**
Stony Brook University

**Stoyan Stoyanov**
Stony Brook University

**Svetlozar Rachev**
Texas Tech University

**Frank J. Fabozzi**
EDHEC Business School

**Abstract**: We extend the classical Cox-Ross-Rubinstein binomial model in two ways. We first develop a binomial model with time-dependent parameters that equate all moments of the pricing tree increments with the corresponding moments of the increments of the limiting Itô price process. Second, we introduce a new trinomial model in the natural (historical) world, again fitting all moments of the pricing tree increments to the corresponding geometric Brownian motion. We introduce the risk-neutral trinomial tree and derive a hedging strategy based on an additional perpetual derivative used as a second asset for hedging in any node of the trinomial pricing tree.

## 1. Introduction

Since its inception, the celebrated binomial pricing model introduced by Cox, Ross, and Rubinstein (1979) (CRR-model hereafter) has been widely used in option pricing theory. There have been several extensions of the CRR-model that have been proposed in the literature. The first are alternative binomial models to ensure faster convergence to the corresponding limiting continuous-time pricing model.[1] The second are trinomial and multinomial models.[2]

In this paper, we address these two extensions, making the following contributions to both of them. More specifically, in the case of the first extension, the basic CRR-pricing tree model is restricted to two limiting price processes – GBM and geometric Poisson process. We show that CRR-model can be extended to a new version with time-dependent parameters. The new CRR-model generates a discrete-pricing model that converges weakly to an Itô process. We then improve the CRR-model with time-dependent parameters, introducing a new binomial model which fits all moments of the price increments to those of the corresponding Itô process.

In the existing literature, the trinomial tree models are defined directly in the risk-neutral world, and in contrast to the CRR-model, it is then not clear which trinomial tree models in the natural world lead to those trinomial tree models in the risk-neutral world. This brings us to the second extension that we investigate in this paper. We first define a trinomial pricing tree fitting all moments to the corresponding limiting geometric Brownian motion in the natural world. We then define the risk-neutral probabilities corresponding to that trinomial tree.

The paper is organized as follows. In Section 2 we introduce binomial pricing trees with time-dependent parameters converging to continuous diffusions. In Section 3, a trinomial tree in the natural world is introduced with all moments fitted to the corresponding geometric Brownian motion (GBM). Our concluding remarks are in Section 4.

---

[1] See Rendleman and Bartter (1979), Jarrow and Rudd (1983, p. 188), Leisen and Reiner (1996), Breen (1991), Walsh (2003), and Kim et al. (2016).

[2] Boyle (1986), Madan, Milne, and Shefrin (1989), Figlewski and Gao (1999), Heston and Zhou (2000), and Lee and Lee (2010).

## 2. Binomial Pricing Trees with Time-Dependent Parameters Converging to Continuous Divisions

In this section, we extend the CRR-model and Kim et al (2016) binomial pricing model (KSRF hereafter), allowing for the coefficients of the model to be time dependent.

### 2.1 CRR-Model with Time-Dependent Parameters

We start with the following generalization of the CRR-model. Consider a market with one risky asset (a stock, designated as $\mathcal{S}^{(N)}, N \in \mathcal{N} \coloneqq \{1,2,...\}$) and one riskless asset (a bond, designated as $\mathcal{B}^{(N)}$). We first introduce the following additional notation. Denote by $\mathrm{E}^{(t)} = \left(\varepsilon^{(\Delta t)}, \varepsilon^{(2\Delta t)}, ..., \varepsilon^{(t)}\right)$ a vector of independent and identically distributed random signs $\varepsilon^{(t)}, t = n\Delta t, n \in \mathcal{N}$. The vector $\mathrm{E}^{(t)}$ defines a stochastic basis[3]

$$\left(\Omega = [0,1], \mathbb{F}^{(N)} = \begin{pmatrix} \{\mathcal{F}^{(t)} = \sigma\left(\mathrm{E}^{(t)}\right)\}, \\ t = n\Delta t, n = 1, ..., N, , \mathcal{F}^{(0)} = \{\emptyset, \Omega\} \end{pmatrix}_{N \in \mathcal{N}}, \mathbb{P} = Leb[0,1]\right).$$

The probability distribution function of $\varepsilon^{(t)}$ is given by $\mathbb{P}(\varepsilon^{(t)} = 1) = 1 - \mathbb{P}(\varepsilon^{(t)} = -1) = p(t, \Delta t)$ where

$$p(t, \Delta t) = \frac{1}{2} + \frac{\mu(t,\Delta t) - \frac{\sigma(t,\Delta t)^2}{2}}{2\sigma(t,\Delta t)} \sqrt{\Delta t}, \qquad \mu(t, \Delta t) \in \mathcal{R}, \sigma(t, \Delta t) > 0. \qquad (1)$$

---

[3] See Shiryaev (1999, p. 297).

The functions $\mu(t, \Delta t)$ and $\sigma(t, \Delta t)$ are such that the following limits $\mu(t) = \lim_{\Delta t \downarrow 0} \mu(t, \Delta t)$, $\sigma(t) = \lim_{\Delta t \downarrow 0} \sigma(t, \Delta t)$, $t \in [0, T]$ exist and have continuous uniformly bounded on $[0, T]$ first derivatives. Finally, we introduce the following function

$$\mathbb{h}(\varepsilon^{(t)}, t, T) = \begin{cases} U(t, \Delta t) \in (1, \infty) \ if \ \varepsilon^{(t)} = 1 \\ D(t, \Delta t) \in (0,1) \ if \ \varepsilon^{(t)} = -1, \end{cases} \quad (2)$$

where

$$U(t, \Delta t) = e^{\sigma(t, \Delta t)\sqrt{\Delta t}}, \ D(t, \Delta t) = e^{-\sigma(t, \Delta t)\sqrt{\Delta t}} \quad (3)$$

In discrete time, the price of $\mathcal{S}^{(N)}$ is some function of the vector of random signs and time, $S^{(N)}(\mathrm{E}^{(t)}, t), t = n\Delta t, \Delta t > 0, n = 0,1, \ldots,$ At the initial time $t = 0$, $\mathrm{E}^{(0)} := \varepsilon^{(0)} = 0$, $S^{(N)}(\mathrm{E}^{(0)}, 0) = S(0) = x^{(0)} > 0$. Assume that the price dynamics is given by

$$S^{(N)}(\mathrm{E}^{(t+\Delta t)}, t + \Delta t) = S^{(N)}(\mathrm{E}^{(t)}, t) \mathbb{h}(\varepsilon^{(t+\Delta t)}, t + \Delta t), t = 0, \ldots T - \Delta t. \quad (4)$$

where $T = N\Delta t < \infty$ is the trading horizon.

Clearly it follows from (2) that the dynamics in (4) defines a binomial tree process. The function defined in (2) describes the proportional increase or decrease of the stock price for any given price movement; the conditions in (3) guarantee that the tree is recombining. Finally, we are interested in the limiting behavior of (4) as the time increment converges to zero and for this reason we assume that $\Delta t$ is sufficiently small, $\Delta t \downarrow 0$, and therefore all terms of order $o(\Delta t)$ are disregarded.

The riskless asset (bond) price dynamics $\beta(t, \Delta t), t = n\Delta t, n \in \mathcal{N}$, is given by

$$\beta(t, \Delta t) = e^{r(t,\Delta t)} = 1 + r(t, \Delta t)\Delta t, r(t, \Delta t) > 0, \qquad (5)$$

where $r(t, \Delta t) \in (0, \mu(t, \Delta t))$ is the riskless rate at $t$. We assume that the limit $r(t) = \lim_{\Delta t \downarrow 0} r(t, \Delta t)$ exists, has continuous uniformly bounded first derivative on $[0, T]$, and $\sup_{t \in [0,T]} \{r(t) + \frac{1}{r(t)}\} < \infty$.

From (3), the log-return

$$R(E^{(t+\Delta t)}, t + \Delta t) = \ln \frac{S(E^{(t+\Delta t)}, t+\Delta t)}{S(E^{(t)}, t)} = \begin{cases} \sigma(t + \Delta t, \Delta t)\sqrt{\Delta t}, & \text{if } \varepsilon^{(t+\Delta t)} = 1 \\ -\sigma(t + \Delta t, \Delta t)\sqrt{\Delta t}, & \text{if } \varepsilon^{(t+\Delta t)} = -1 \end{cases} \qquad (6)$$

has conditional first moment $\mathbb{E}_t R(E^{(t+\Delta t)}, t + \Delta t) = \left(\mu(t + \Delta t, \Delta t) - \frac{\sigma(t+\Delta t,\Delta t)^2}{2}\right)\Delta t$, and conditional variance $\mathrm{var}_t R(E^{(t+\Delta t)}, t + \Delta t) = \sigma(t + \Delta t, \Delta t)^2 \Delta t$. By Proposition 3 in Davydov and Rotar (2008), if $R^{(N)}(s), 0 \leq s \leq T$, is defined as a piecewise random process (a random polygon) connecting with linear segments the vertexes $\left(n\Delta t, R(E^{(n\Delta t)}, n\Delta t)\right), n = 0, \ldots, N$, then, the process $\left(R^{(N)}(s)\right)_{0 \leq s \leq T}$ converges weakly in $C\{0, T\}$ to Gaussian process $(R(s))_{0 \leq s \leq T}$. The limiting pricing process $R(s), 0 \leq s \leq T$ is an Itô process (a continuous diffusion):

$$R(s) = \int_0^s \left(\mu(u) - \frac{\sigma(u)^2}{2}\right) du + \int_0^s \sigma(u)\, dB(u), 0 \leq s \leq T,$$

where $B(s), 0 \leq s \leq T$, is a standard Brownian motion generating the canonical stochastic basis $\left(\Omega, \mathbb{F} = \left(\{\mathcal{F}^{(s)} = \sigma(B(u), 0 \leq u \leq s)\}, s \in [0, T]\right)_{T>0}, \mathbb{P}\right)$. Given the initial stock price $S(0) > 0$, and defining the discrete-price process as $\left(S^{(N)}(s) = S(0)e^{R^{(N)}(s)}\right)_{0 \leq s \leq T}$, we have that $\left(S^{(N)}(s)\right)_{0 \leq s \leq T}$ weakly converges in $C\{0, T\}$ to the Itô price process $S(s) = S(0)e^{R(s)}, s \geq 0$, with

$$dS(s) = \mu(s)S(s)ds + \sigma(s)S(s)dB(s), 0 \leq s \leq T. \qquad (7)$$

Thus, (7) represents the continuous-price dynamics of the stock $\mathcal{S}$, while

$$\beta(s) = \beta(0) e^{\int_0^s r(u)du}, 0 \le s \le T, \tag{8}$$

represents the continuous-price dynamics of a riskless bond $\mathcal{B}$.

We now turn our attention to a risk-neutral binomial tree corresponding to the binomial tree (see (4)) in the natural world defined by the triplet $(\Omega, \mathbb{F}^{(N)}, \mathbb{P})$. To this end, we consider a derivative $\mathcal{G}^{(N)}$ with price process $G^{(N)}(\mathrm{E}^{(t)}, t), t = n\Delta t, n = 0, \ldots, N,$ on $(\Omega, \mathbb{F}^{(N)}, \mathbb{P})$, given by

$$G^{(N)}\big(\mathrm{E}^{(t+\Delta t)}, t+\Delta t\big) = \begin{cases} G^{(N,+)}\big(\mathrm{E}^{(t)}, t\big) \text{ if } \varepsilon^{(t+1)} = 1 \\ G^{(N,-)}\big(\mathrm{E}^{(t)}, t\big) \text{ if } \varepsilon^{(t+1)} = -1, \end{cases} \quad t = 0, \ldots, T-\Delta t,$$

and an initial value $G^{(N)}(\mathrm{E}^{(0)}, 0) = G^{(N)}(0)$. At the terminal time $T = N\Delta t$, the derivative value is given by $G^{(N)}(T) = g(S(T))$ for some payoff function $g(x), x > 0$. Suppose a trader takes a short position in $\mathcal{G}^{(N)}$. At $t = n\Delta t$, the trader forms a portfolio $\mathcal{P}^{(N)}(t), t = n\Delta t,$ with a price process $P^{(N)}(\mathrm{E}^{(t)}, t) = -G^{(N)}(\mathrm{E}^{(t)}, t) + \Psi(\mathrm{E}^{(t)}, t) S^{(N)}(\mathrm{E}^{(t)}, t)$. The trader determines $\Psi(\mathrm{E}^{(t)}, t)$ so that at $t + \Delta t$, portfolio $\mathcal{P}^{(N)}(t + \Delta t)$ is riskless, that is

$$P^{(N)}\big(\mathrm{E}^{(t+\Delta t)}, t+\Delta t\big) = -G^{(N,+)}\big(\mathrm{E}^{(t)}, t\big) + \Psi\big(\mathrm{E}^{(t)}, t\big) S^{(N)}\big(\mathrm{E}^{(t)}, t\big) e^{\sigma(t+\Delta t, \Delta t)\sqrt{\Delta t}}$$

$$= -G^{(N,-)}\big(\mathrm{E}^{(t)}, t\big) + \Psi\big(\mathrm{E}^{(t)}, t\big) S^{(N)}\big(\mathrm{E}^{(t)}, t\big) e^{-\sigma(t+\Delta t, \Delta t)\sqrt{\Delta t}}$$

Thus, $\Psi(\mathrm{E}^{(t)}, t) = \frac{G^{(N,+)}(\mathrm{E}^{(t)},t) - G^{(N,-)}(\mathrm{E}^{(t)},t)}{2S^{(N)}(\mathrm{E}^{(t)},t)\sigma(t+\Delta t, \Delta t)\sqrt{\Delta t}}$. Then, $P^{(N)}(\mathrm{E}^{(t)}, t) = e^{-r(t+\Delta t, \Delta t)\Delta t} P^{(N)}(\mathrm{E}^{(t+\Delta t)}, t + \Delta t)$ and thus,

$$G^{(N)}(\mathrm{E}^{(t)}, t) = e^{-r(t+\Delta t, \Delta t)} \begin{Bmatrix} q(t + \Delta t, \Delta t) G^{(N,+)}(\mathrm{E}^{(t)}, t) + \\ +(1 - q(t + \Delta t, \Delta t)) G^{(N,-)}(\mathrm{E}^{(t)}, t) \end{Bmatrix}, \tag{9}$$

where

$$q(t + \Delta t, \Delta t) := \frac{1}{2} + \frac{r(t+\Delta t,\Delta t) - \frac{\sigma(t+\Delta t,\Delta t)^2}{2}}{2\sigma(t+\Delta t,\Delta t)}\sqrt{\Delta t} \qquad (10)$$

and $1 - q(t + \Delta t, \Delta t)$ are the risk-neutral probabilities. Now, the binomial price process in the risk-neutral world is given by

$$S^{(N)}\left(\mathrm{E}^{(\mathbb{Q},t+\Delta t)}, t + \Delta t\right) = S^{(N)}\left(\mathrm{E}^{(\mathbb{Q},t)}, t\right)\mathbb{h}\left(\varepsilon^{(\mathbb{Q},t+\Delta t)}, t + \Delta t\right) \qquad (11)$$

where $\mathrm{E}^{(\mathbb{Q},t)} = \left(\varepsilon^{(\mathbb{Q},\Delta t)}, \varepsilon^{(\mathbb{Q},2\Delta t)}, \ldots, \varepsilon^{(\mathbb{Q},t)}\right)$, and $\varepsilon^{(\mathbb{Q},t)}, t = n\Delta t, n \in \mathcal{N}$ are iid random sign $\mathbb{P}(\varepsilon^{(\mathbb{Q},t)} = 1) = 1 - \mathbb{P}(\varepsilon^{(\mathbb{Q},t)} = -1) = q(t,\Delta t)$. As with (4) and (7), in the limiting case $\Delta t \downarrow 0$, $\left(S^{(N)}\left(\mathrm{E}^{(\mathbb{Q},t+\Delta t)}, t + \Delta t\right)\right)_{t=n\Delta t, n=0,1,\ldots,N-1}$ generates a polygonal process in $C[0,T]$ weakly converging to an Itô process $\left(S^{(\mathbb{Q})}(s)\right)_{0 \leq s \leq T}$, given by

$$dS^{(\mathbb{Q})}(s) = r(s)S^{(\mathbb{Q})}(s)ds + \sigma(s)S^{(\mathbb{Q})}(s)dB^{(\mathbb{Q})}(s), 0 \leq s \leq T, S^{(\mathbb{Q})}(s) = S(0), \qquad (12)$$

where $B^{(\mathbb{Q})}(s), 0 \leq s \leq T$, is a Brownian motion on $(\Omega, \mathbb{F}, \mathbb{Q}), \mathbb{Q} \sim \mathbb{P}$, which on $\mathbb{P}$ is given by $B^{(\mathbb{Q})}(s) = B(s) + \int_0^s \theta(u)du$, with $\theta(s) = \frac{\mu(s) - r(s)}{\sigma(s)}, 0 \leq s \leq T$ being the market price of risk. We thus extended the CRR model in the case when the model parameters are time dependent.

### 2.2. KSRF-Model with Time-Dependent Parameters

A drawback of the CCR-model is the fact that the risk-neutral probability $q(t, \Delta t)$ is independent of the initial probability $p(t, \Delta t)$. Thus, even if $p(t, \Delta t) \uparrow 1$ or $p(t, \Delta t) \downarrow 0$, the risk-neutral probability $q(t, \Delta t)$ is unchanged. This leads to an unsettling discontinuity of the option price at $p(t, \Delta t) = 0$ and $p(t, \Delta t) = 1$. The KSRF-model resolved this issue, generalizing the CRR-model, so that the risk-neural probability becomes a continuous function of $p(t, \Delta t)$.

We now consider the following generalization of KSRF-model. Assume that $S^{(N)}$-priced dynamics $S^{(N)}(\mathrm{E}^{(t)}, t), t = n\Delta t, \Delta t > 0, n = 0,1,\ldots,N, \mathrm{E}^{(0)} := \varepsilon^{(0)} = 0, S^{(N)}(\mathrm{E}^{(0)}, 0) = S(0) = x^{(0)} > 0$ is given by the following binomial tree

$$S^{(N)}(\mathrm{E}^{(t+\Delta t)}, t + \Delta t) = S^{(N)}(\mathrm{E}^{(t)}, t)\mathrm{g}(\varepsilon^{(t+\Delta t)}, t + \Delta t), t = 0, \ldots T - \Delta t. \qquad (13)$$

In (13),

**KSRF1:** $\quad \mathrm{g}(\varepsilon^{(t)}, t, T) = \begin{cases} 1 + \mu(t, \Delta t)\Delta t + \sqrt{\frac{1-p(t,\Delta t)}{p(t,\Delta t)}} \sigma(t, \Delta t)\sqrt{\Delta t} & if \quad \varepsilon^{(t)} = 1 \\ 1 + \mu(t, \Delta t)\Delta t - \sqrt{\frac{p(t,\Delta t)}{1-p(t,\Delta t)}} \sigma(t, \Delta t)\sqrt{\Delta t} & if \quad \varepsilon^{(t)} = -1 \end{cases} \qquad (14)$

**KSRF2:** $\quad \mu(t, \Delta t)$ and $\sigma(t, \Delta t)$ are defined as in (1);

**KSRF3:** $\quad \mathrm{E}^{(t)} = (\varepsilon^{(\Delta t)}, \varepsilon^{(2\Delta t)}, \ldots, \varepsilon^{(t)})$, and $\varepsilon^{(t)}, t = n\Delta t, n \in \mathcal{N}$ $\varepsilon^{(t)}, t = n\Delta t$ are iid random signs $\varepsilon^{(t)}$ $\mathbb{P}(\varepsilon^{(t)} = 1) = 1 - \mathbb{P}(\varepsilon^{(t)} = -1) = p(t, \Delta t) \in (0,1)$;

**KSRF4:** The limiting function $p(t) = \lim_{\Delta t \downarrow 0} p(t, \Delta t)$ exists and have continuous uniformly bounded on $[0, T]$ first derivative.

First, let us verify that the tree given by (13) is a recombining tree. From $p(t + \Delta t, \Delta t) = p(t, \Delta t) + \frac{\partial p(t,\Delta t)}{\partial t} \Delta t$ and $\sigma(t + \Delta t, \Delta t) = \sigma(t, \Delta t) + \frac{\partial \sigma(t,\Delta t)}{\partial t} \Delta t$, and recalling that all terms of order $o(\Delta t)$ are disregarded, we have that

$\left(1 + \mu(t, \Delta t)\Delta t + \sqrt{\frac{1-p(t,\Delta t)}{p(t,\Delta t)}} \sigma(t, \Delta t)\sqrt{\Delta t}\right)\left(1 + \mu(t + \Delta t, \Delta t)\Delta t - \sqrt{\frac{p(t+\Delta t,\Delta t)}{1-p(t+\Delta t,\Delta t)}} \sigma(t + \Delta t, \Delta t)\sqrt{\Delta t}\right) = \left(1 + \mu(t, \Delta t)\Delta t - \sqrt{\frac{p(t,\Delta t)}{1-p(t,\Delta t)}} \sigma(t, \Delta t)\sqrt{\Delta t}\right)\left(1 + \mu(t + \Delta t, \Delta t)\Delta t + \sqrt{\frac{1-p(t+\Delta t,\Delta t)}{p(t+\Delta t,\Delta t)}} \sigma(t + \Delta t, \Delta t)\sqrt{\Delta t}\right),$

showing that the tree given by (12) is a recombining tree.

Let us show next that the trees given by (13) and (14) have all moments of its $\Delta t$-increments equal to the corresponding moments of the $\Delta t$-increments of the Itô process

$$S(t) = S(0)e^{\int_0^t \left(\mu(u) - \frac{\sigma(u)^2}{2}\right)du + \int_0^t \sigma(u)dB(u)}, t \in [0, T]. \tag{15}$$

From (13), (14) and (15), for $t = n\Delta t$, and any $\varsigma > 0$, and ignoring the terms of order $o(\Delta t)$, it follows that

$$\mathbb{E}_t\left(\frac{S^{(N)}(\mathrm{E}^{(t+\Delta t)}, t+\Delta t)}{S(\mathrm{E}^{(t)}, t)}\right)^\varsigma = \mathbb{E}_t\left(\frac{S(t+\Delta t)}{S(t)}\right)^\varsigma \text{ for all } \varsigma > 0. \tag{16}$$

As with (7), in the limiting case $\Delta t \downarrow 0$, $\left(S^{(N)}(\mathrm{E}^{(t+\Delta t)}, t+\Delta t)\right)_{t=n\Delta t, n=0,1,\ldots,N-1}$ generates a polygonal process in $C[0,T]$ weakly converging to an Itô process $(S(s))_{0 \leq s \leq T}$, given by (14).

We now turn our attention to the risk-neutral binomial tree corresponding to (13) and (14). To this end, consider again the derivative $\mathcal{G}^{(N)}$ with price process $G^{(N)}(\mathrm{E}^{(t)}, t), t = n\Delta t, n = 0, \ldots, N,$ on $(\Omega, \mathbb{F}^{(N)}, \mathbb{P})$. Suppose that a trader takes a short position in $\mathcal{G}^{(N)}$. At $t = n\Delta t$, the trader forms a portfolio $\mathcal{P}^{(N)}(t), t = n\Delta t$, with price process $P^{(N)}(\mathrm{E}^{(t)}, t) = -G^{(N)}(\mathrm{E}^{(t)}, t) + \Psi(\mathrm{E}^{(t)}, t)S^{(N)}(\mathrm{E}^{(t)}, t)$. The trader determines $\Psi(\mathrm{E}^{(t)}, t)$ so that at $t + \Delta t$, portfolio $\mathcal{P}^{(N)}(t + \Delta t)$ is riskless. Thus,

$$P^{(N)}(\mathrm{E}^{(t+\Delta t)}, t+\Delta t) = -G^{(N,-)}(\mathrm{E}^{(t)}, t) +$$

$$+\Psi(\mathrm{E}^{(t)}, t)S^{(N)}(\mathrm{E}^{(t)}, t)\left(1 + \mu(t+\Delta t, \Delta t)\Delta t - \sqrt{\frac{p(t+\Delta t, \Delta t)}{1 - p(t+\Delta t, \Delta t)}}\sigma(t+\Delta t, \Delta t)\sqrt{\Delta t}\right)$$

As a result,

$$\Psi(E^{(t)},t) = \frac{G^{(N,+)}(E^{(t)},t) - G^{(N,-)}(E^{(t)},t)}{S^{(N)}(E^{(t)},t)\sigma(t+\Delta t,\Delta t)\sqrt{\Delta t}}\sqrt{(1-p(t,\Delta t))p(t,\Delta t)}$$

Denote by $\theta(t,\Delta t) = \frac{\mu(t,\Delta t) - r(t,\Delta t)}{\sigma(t,\Delta t)}$ the market price of risk at $t = n\Delta t$. Then, from $P^{(N)}(E^{(t)},t) = e^{-r(t+\Delta t,\Delta t)\Delta t}P^{(N)}(E^{(t+\Delta t)},t+\Delta t)$, it follows that

$$G^{(N)}(E^{(t)},t) = e^{-r(t+\Delta t,\Delta t)}\left\{\begin{array}{l}q^{(*)}(t+\Delta t,\Delta t)G^{(N,+)}(E^{(t)},t) + \\ + \left(1 - q^{(*)}(t+\Delta t,\Delta t)\right)G^{(N,-)}(E^{(t)},t)\end{array}\right\}, \quad (17)$$

where

$$q^{(*)}(t+\Delta t,\Delta t) := p(t+\Delta t,\Delta t) - \theta(t+\Delta t,\Delta t)\sqrt{(1-p(t,\Delta t))p(t,\Delta t)}\sqrt{\Delta t} \quad (18)$$

and $1 - q^{(*)}(t+\Delta t,\Delta t)$ are the corresponding risk-neutral probabilities. Notice that $q^{(*)}$ depends continuously on $p(t,\Delta t)$, resolving the discontinuity problem of the CRR-model when $p(t,\Delta t) \downarrow 0$, or $p(t,\Delta t) \uparrow 1$.

Now, the binomial price process in the risk-neutral world is given by

$$S^{(N)}(E^{(*,t+\Delta t)},t+\Delta t) = S^{(N)}(E^{(*,t)},t)\mathbb{g}(\varepsilon^{(*,t+\Delta t)},t+\Delta t) \quad (19)$$

where $E^{(*,t)} = (\varepsilon^{(*,\Delta t)},\varepsilon^{(*,2\Delta t)},\ldots,\varepsilon^{(*,t)})$, and $\varepsilon^{(*,t)}, t = n\Delta t, n \in \mathcal{N}$ are iid random signs $\mathbb{P}(\varepsilon^{(*,t)} = 1) = 1 - \mathbb{P}(\varepsilon^{(*,t)} = -1) = q^{(*)}(t,\Delta t)$. Note that

$$\mathbb{E}_t \frac{S^{(N)}(E^{(*,t+\Delta t)},t+\Delta t)}{S^{(N)}(E^{(*,t)},t)} = 1 + r(t,\Delta t)\Delta t \text{ and } var_t \frac{S^{(N)}(E^{(*,t+\Delta t)},t+\Delta t)}{S^{(N)}(E^{(*,t)},t)} = \sigma(t,\Delta t)^2\Delta t.$$

As with (7), in the limiting case $\Delta t \downarrow 0$, $\left(S^{(N)}(E^{(*,t+\Delta t)},t+\Delta t)\right)_{t=n\Delta t, n=0,1,\ldots,N-1}$ generates a polygonal process in $C[0,T]$ weakly converging to an Itô process $\left(S^{(\mathbb{Q})}(s)\right)_{0\leq s\leq T}$, given by (12).

## 3. New Trinomial Tree with all Moments Fitted to the Corresponding GBM

In this section, we extend the previous section results to trinomial pricing tree models. For simplicity of the exposition we shall consider the case with constant coefficients. The general case follows the same arguments as in Section 2.

The classical trinomial pricing model[4] is directly defined in the risk-neutral world and has the form

$$S\left(\mathrm{E}^{(TR,t+\Delta t)}, t + \Delta t\right) = S\left(\mathrm{E}^{(TR,t)}, t\right) \begin{cases} 1 + \frac{3}{2}\sigma^2 \Delta t + \sigma\sqrt{3\Delta t}, & \text{if } \varepsilon^{(TR,t+\Delta t,\Delta t)} = 1, \\ 1, & \text{if } \varepsilon^{(TR,t+\Delta t,\Delta t)} = 0, \\ 1 + \frac{3}{2}\sigma^2 \Delta t - \sigma\sqrt{3\Delta t}, & \text{if } \varepsilon^{(TR,t+\Delta t,\Delta t)} = -1. \end{cases} \quad (20)$$

In (20), $S\left(\mathrm{E}^{(TR,0)}, 0\right) = S(0), t = n\Delta t, T = N\Delta t,$ and $\mathrm{E}^{(TR,n\Delta t)} = \left(\varepsilon^{(TR,\Delta t,\Delta t)}, \ldots, \varepsilon^{(TR,n\Delta t,\Delta t)}\right)$, where $\varepsilon^{(TR,n\Delta t,\Delta t)}, n = 1, \ldots, N,$ are iid random variables with $\mathbb{P}\left(\varepsilon^{(TR,t,\Delta t)} = 1\right) := q^{(u)} = \frac{1}{6} + \sqrt{\frac{\Delta t}{12\sigma^2}}\left(r - \frac{\sigma^2}{2}\right), \mathbb{P}\left(\varepsilon^{(TR,t,\Delta t)} = 0\right) := q^{(n)} = \frac{2}{3}, \mathbb{P}\left(\varepsilon^{(TR,t,\Delta t)} = -1\right) := q^{(d)} = \frac{1}{6} - \sqrt{\frac{\Delta t}{12\sigma^2}}\left(r - \frac{\sigma^2}{2}\right), q^{(u)} + q^{(n)} + q^{(d)} = 1.$ The parameters $q^{(u)}, q^{(n)}$ and $q^{(d)}$ are simply calculated by fitting the mean and the variance of $S\left(\mathrm{E}^{(TR,\Delta t)}, \Delta t\right)$ to the mean and the variance of $S^{\mathbb{Q}}(\Delta t)$ (eliminating all terms of order $o(\Delta t)$), where

$$S^{\mathbb{Q}}(t) = S(0) e^{\left(r - \frac{\sigma^2}{2}\right)t + \sigma B^{\mathbb{Q}}(t)}, t \geq 0. \quad (21)$$

In (21), $\mathbb{Q} \sim \mathbb{P}$ and $B^{\mathbb{Q}}(t), t \geq 0,$ is a Brownian motion on $\mathbb{Q}$, while on $\mathbb{P}$ it has the representation $B^{\mathbb{Q}}(t) = B(t) + \theta t,$ where $\theta = \frac{\mu - r}{\sigma}$ is the market price of risk.

---

[4] See for example Hull (2012, p 444).

We now introduce a new trinomial tree matching all moments of $S(\mathrm{E}^{(TR,\Delta t)}, \Delta t)$ and $S^{\mathbb{Q}}(\Delta t)$. Our trinomial model has the following form:

$$S(\mathrm{E}^{(U,t+\Delta t)}, t + \Delta t) = S(\mathrm{E}^{(U,t)}, t) \begin{cases} 1 + \left(\mu + \frac{\sigma^2}{4}\right)\Delta t + \sqrt{\frac{3}{2}}\sigma\sqrt{\Delta t}, & if\ \varepsilon^{(U,t+\Delta t,\Delta t)} = 1 \\ 1 + \left(\mu - \frac{\sigma^2}{2}\right)\Delta t, & if\ \varepsilon^{(U,t+\Delta t,\Delta t)} = 0 \\ 1 + \left(\mu + \frac{\sigma^2}{4}\right)\Delta t - \sqrt{\frac{3}{2}}\sigma\sqrt{\Delta t}, & if\ \varepsilon^{(U,t+\Delta t,\Delta t)} = -1. \end{cases} \quad (22)$$

In (22), $S(\mathrm{E}^{(U,0)}, 0) = S(0), t = n\Delta t, T = N\Delta t$, and $\mathrm{E}^{(U,n\Delta t)} = \left(\varepsilon^{(TR,\Delta t,\Delta t)}, \dots, \varepsilon^{(TR,n\Delta t,\Delta t)}\right)$, where $\varepsilon^{(TR,n\Delta t,\Delta t)}, n = 1, \dots, N$, are iid random variables with $\mathbb{P}(\varepsilon^{(TR,t,\Delta t)} = 1) := p^{(u)} = \mathbb{P}(\varepsilon^{(TR,t,\Delta t)} = 0) := p^{(n)} = \mathbb{P}(\varepsilon^{(TR,t,\Delta t)} = -1) := p^{(d)} = \frac{1}{3}$. Indeed, the tree given by (22) is a recombining tree, as

$$\left(1 + \left(\mu + \frac{\sigma^2}{4}\right)\Delta t + \sqrt{\frac{3}{2}}\sigma\sqrt{\Delta t}\right)\left(1 + \left(\mu + \frac{\sigma^2}{4}\right)\Delta t - \sqrt{\frac{3}{2}}\sigma\sqrt{\Delta t}\right) = \left(1 + \left(\mu - \frac{\sigma^2}{2}\right)\Delta t\right)^2$$

Next for all $\varsigma > 0$, as $\Delta t \downarrow 0$, $\mathbb{E}\left(\frac{S(\mathrm{E}^{(U,\Delta t)}, \Delta t)}{S(0)}\right)^{\varsigma} = 1 + \varsigma\left(\mu + \frac{\varsigma - 1}{2}\sigma^2\right)\Delta t = \mathbb{E}\left(\frac{S(\Delta t)}{S(0)}\right)^{\varsigma}$,

where $S(t), t \geq 0$, is the GBM given by

$$S(t) = S(0)e^{\left(\mu - \frac{\sigma^2}{2}\right)t + \sigma B(t)}, t \geq 0, S(0) > 0 \quad (23)$$

That is, we show the remarkable property that all moments of the $\Delta t$ - increments of the trinomial tree (22) and the GBM (20) coincide when the terms of order $o(\Delta t)$ are disregarded. This leads to the weak convergence of the $D[0, T]$-process generated by (22) to the GBM (23).

Now, the corresponding risk-neutral tree has the form

$$S^{\mathbb{Q}}\left(\mathrm{E}^{(U,t+\Delta t)}, t+\Delta t\right) =$$

$$S^{\mathbb{Q}}\left(\mathrm{E}^{(U,t)}, t\right) \begin{cases} 1+\left(r+\frac{\sigma^2}{4}\right)\Delta t + \sqrt{\frac{3}{2}}\sigma\sqrt{\Delta t}, \ w.p.\ q^{(u)} = \frac{1}{3} \\ 1+\left(r-\frac{\sigma^2}{2}\right)\Delta t, 1, w.p.\ q^{(n)} = \frac{1}{3} \\ 1+\left(r+\frac{\sigma^2}{4}\right)\Delta t - \sqrt{\frac{3}{2}}\sigma\sqrt{\Delta t}\ w.p.\ q^{(d)} = \frac{1}{3}. \end{cases} \quad (24)$$

Then, the $\mathbb{E}\left(\frac{S^{\mathbb{Q}}\left(\mathrm{E}^{(U,\Delta t)}, \Delta t\right)}{S(0)}\right)^k = \mathbb{E}\left(\frac{S^{\mathbb{Q}}(\Delta t)}{S(0)}\right)^k, k = 1,2$, where $S^{\mathbb{Q}}(t), t \geq 0$, is the risk-neutral stock price dynamics given by (21). Thus, we have defined a new trinomial recombining tree model given by (24) fitting all moments $\mathbb{E}\left(\frac{S^{\mathbb{Q}}\left(\mathrm{E}^{(U,\Delta t)}, \Delta t\right)}{S(0)}\right)^k$ to the corresponding moments, $\mathbb{E}\left(\frac{S^{\mathbb{Q}}(\Delta t)}{S(0)}\right)^k, k = 1,2$, of the continuous risk-neutral GBM given by (21).

## 4. Conclusion

In this paper, we extend the CRR-model in two directions. First, we define a binomial pricing model with time-dependent parameters, fitting all moments of the increments of the binomial pricing tree to the increments of the limiting Itô pricing process. Second, we introduce a trinomial pricing model with all moment of the increments of the pricing tree fitted to the corresponding increments of the geometric Brownian motion.